\documentclass[aps,prl, reprint,floatfix,superscriptaddress,showpacs,amsfonts
,amssymb,amsmath,preprintnumbers]{revtex4-1}
\usepackage{bm}
\usepackage{epsfig}
\usepackage{amsmath}
\usepackage{amssymb}
\usepackage{graphicx}
\usepackage{bm}
\usepackage{color}
\usepackage{subfigure}

\newcommand{\fig}[1]{Fig.~\ref{#1}}

\newcommand{\eq}[1]{Eq.~(\ref{#1})}
\newcommand{\eqs}[2]{Eqs.~(\ref{#1}--\ref{#2})}

\newcommand{\be}{\begin{equation}}
\newcommand{\ee}{\end{equation}}

\newcommand\bea{\begin{eqnarray}}
\newcommand\eea{\end{eqnarray}}

\begin{document}
\setcounter{secnumdepth}{1}

\title{The effects of collisions on the generation and suppression of temperature anisotropies and the Weibel instability}

\author{K. M. Schoeffler}
\affiliation{GoLP/Instituto de Plasmas e Fus\~ao Nuclear,
Instituto Superior T\'ecnico,\\ Universidade de Lisboa, 1049-001 Lisboa, Portugal}
\author{L. O. Silva}
\affiliation{GoLP/Instituto de Plasmas e Fus\~ao Nuclear,
Instituto Superior T\'ecnico,\\ Universidade de Lisboa, 1049-001 Lisboa, Portugal}

\date{\today}

\begin{abstract}
	The expansion of plasma with non-parallel temperature and density
	gradients, and the generation of magnetic field via the Biermann
	battery is modeled using particle-in-cell simulations that include
	collisional effects via Monte Carlo methods. A scaling of the degree of
	collisionality shows that an anisotropy can be produced, and drive the
	Weibel instability, for gradient scales shorter than the mean free
	path.  For larger collision rates, the Biermann battery dominates as
	the cause of magnetic field generation. When the most energetic
	particles remain collisionless, the Nernst effect causes the Biermann
	field to be dragged with the heat flux, piled up, and enhanced.
\end{abstract}


\maketitle

%***********************************************************************************
\section{Introduction}
%Introduce magnetic field generation by laser interaction in context of
%Biermann battery, and the kinetic effect of the Weibel instability
%which is present in collisionless systems.
Identifying the mechanisms responsible for the generation of various magnetic
fields present throughout the universe is a major topic of study in
astrophysics~\citep{Kulsrud2008}. Two common candidates are the Biermann
battery~\citep{Biermann1950,Stamper71}, which is driven by violent interactions
with unmagnetized plasmas that leave the temperature and density gradients
misaligned, and the Weibel instability~\citep{Weibel1959}, driven by
temperature anisotropies. The Weibel instability in particular is only possible
in collisionless systems, often found in astrophysics due to high temperatures
and low densities of plasmas in space. In laser-plasma interaction experiments
on earth, both the Biermann battery~\citep{Stamper71,Nilson06,Li07,Kugland2012}
and the Weibel
instability~\citep{Kugland2012,Huntington2015,Gode2017,Ngirmang2019arXiv} play
an important role. The sudden heating of the plasma by a laser leads to the
gradients required for the Biermann battery, and although the densities can be
large, the temperatures can be sufficiently high that the plasma is
collisionless, can become anisotropic, and thus become unstable to the Weibel
instability.

%Pose question. What is the limit to the collisionality where this Weibel instability occurs.
A natural question is; what level of collisionality is required for the Weibel
instability to be suppressed? It was shown in
Refs.~\citep{Schoeffler2014,Schoeffler2016} that for a collisionless system,
the Weibel instability is the dominant magnetic field for sufficiently large
gradient scales, $L/d_e > 100$, where $L$ is the temperature or density
gradient length scale, $d_e = c/\omega_{pe}$ is the electron skin depth,
$\omega_{pe} = \sqrt{4\pi n_e e^2/m_e}$ is the plasma frequency, $m_e$, $e$,
$n_e$ are the respective electron mass, charge, and density, and $c$ is the
speed of light. However, this is no longer the case for a sufficiently
collisional plasma (e.g. Ref.~\cite{Fox2018}), where the Biermann field becomes
the dominant field. Here we show, using PIC simulations, what level of
collisionality is required for the Biermann field to dominate over the Weibel
instability.

%Theory of Weibel instability in collisional systems. It is suppressed for large collision
%rates.
Although the original Weibel formulation assumes a collisionless plasma, the
instability has been formulated for a semi-collisional system showing a
dependence on the electron collision rate $\nu_e$~\citep{Wallace1987, Wallace1991}.
\begin{equation}
	\label{collisionalWeibel}
	\gamma_W = \gamma_{W0} - \frac{A}{1+A} \nu_e 
\end{equation}
Here $\gamma_{W0}$ is the collisionless growth rate of the Weibel instability~\cite{Weibel1959},
which depends on the perpendicular electron temperature $T_{e\perp}$,
$\omega_{pe}$, and the temperature anisotropy $A=T_{e\parallel}/T_{e\perp}-1$,
where $T_{e\parallel}$ and $T_{e\perp}$ are the two temperatures in a
bi-Maxwellian distribution. Here, parallel is defined by the direction that has
a different temperature, and we have assumed $T_{e\parallel} > T_{e\perp}$. 
The collision rate is
$\nu_e = \nu_0\left(1 + Z\right)$ where
\be
\nu_0 = \sqrt{\frac{1}{m_e}}\frac{4 \pi n_e e^4}{T_e^{3/2}} 
\ln \Lambda_C,
\ee
$T_e \equiv (2 T_{e\perp} + T_{e\parallel})/3$ is the
electron temperature, $\ln \Lambda_C$ is
the Coulomb logarithm, and $Z$ is the degree of ionization.  $\nu_e$ can be
divided into electron-electron collisions $\nu_{ee} = \nu_0$, and electron-ion
collisions $\nu_{ei} = \nu_0 Z$.

%Explain that the instability is not suppressed by this collisional effect, but rather that
%the collisions prevent the generation of anisotropies that drive the instability.
It turns out, however, that the modifications to $\gamma_W$ due to collisions
are not relevant in most regimes.  The growth rate drops to zero even when
these modifications are negligible, because the instability is driven by $A$
and collisions cause $A$ to decay.  The ratio of the collisional term in
\eq{collisionalWeibel} to $\gamma_{W0}$ is $\sim \nu_e / (\omega_p v_T /c)=
(d_e/L_T) \nu_e L_T/v_T$, where $L_T\equiv T_e/dT_e/dx$ is the temperature
gradient scale, and $v_T=\sqrt{T_e/m_e}$ is the thermal velocity.  As long as
$L_T \gg d_e$, the collision term can be neglected when $\nu_e L_T/v_T \sim 1$,
which we will show is when the Weibel instability is suppressed.

The growth rate drops to zero when the time scale of the anisotropy generation
$t_{Ag}$ reaches the time scale of collisional relaxation time of the
anisotropy $t_{Ar}$.  Ref.~\citep{Schoeffler2018} showed that a gradient in an
isotropic Maxwellian temperature leads to a temperature anisotropy $A$,
saturating at a time scale $t_{Ag} \approx L_T/v_T$ with a value $A \sim 1$.
Refs. \citep{Hellinger2009,Hellinger2010} showed that the relaxation rate is
\be
\label{Adecayrate}
\nu_{A} \equiv \frac{1}{A}\frac{dA}{dt} = - \nu_T\left(A+3\right),
\ee
where
\be
\label{Tdecayrate}
%\nu_{T} = \frac{8 \sqrt{\pi} n_e e^4}{15 \sqrt{m_e} T_{e\parallel}^{3/2}} 
\nu_{T} = \frac{2\nu_0(T_{e\perp})}{15 \sqrt{\pi}} 
\left(1+\sqrt{2}Z\right)
F_{2 \frac{3}{2} \frac{7}{2}}(A/(A+1)), 
\ee
$\nu_0(T_{e\perp})$ is $\nu_0$ replacing $T_e$ with $T_{e\perp}$,
$F_{2 \frac{3}{2} \frac{7}{2}}(x) 
= 15/4 \left[-3 + \left(x+3\right)\phi(x)\right] x^{-2} $ 
is the Gaussian Hypergeometric function, and 
\be
    \phi(x)=
\begin{cases}
	\frac{\tan^{-1}\left(\sqrt{x}\right)}{\sqrt{x}} ,&~\text{if } x > 0\\
    1 ,& \text{if } x = 0\\
	\frac{\tanh^{-1}\left(\sqrt{-x}\right)}{\sqrt{-x}} ,&~\text{if } x < 0.
\end{cases}
\ee
The factors of $1$ and $\sqrt{2}Z$ are a result of the respective
electron-electron and electron-ion collisions.  One finds $\nu_A \sim \nu_e$
using \eqs{Adecayrate}{Tdecayrate}, since $F_{2 \frac{3}{2} \frac{7}{2}}(x)$ is
a constant of order unity (approaching $1$ for small $A$) as long as $A
\lesssim 1$. Remember the anisotropy is expected to reach a maximum of $A \sim
1$.  The Weibel instability is thus suppressed if
\be
\label{Weibelsuppression}
\frac{t_{Ag}}{t_{Ar}} = \frac{\nu_A L_T}{v_T} \sim \frac{\nu_e L_T}{v_T} > 1
\ee
i.e. if the gradient scale is bigger than the mean free path of an electron.

Even if the Weibel instability is fully suppressed, it has been suggested that
an instability known as the thermomagnetic instability~\citep{Tidman1974} may
also generate filamentary magnetic fields due to a parallel density and
temperature gradient due to a combination of the Righi-Leduc and Biermann
battery effects, for collisional systems. However, Ref.~\cite{Sherlock2020}
showed that this instability is suppressed due to the Nernst effect, and that
the growth of the Biermann battery is reduced. The Nernst effect, when only the
most energetic electrons are frozen into the magnetic field, is expected to
drag Biermann generated fields in the direction of heat flux allowing them to
pile
up~\cite{Nishiguchi1984,Kho1985,Ridgers2008,Willingale2010,Willingale2010b,Joglekar2014,Joglekar2016}.

%For $t_{Ag} \ll t_{Ar}$, the temperature gradient causes the anisotropy to grow
%to close to unity at a timescale as $L_T/v_T$.  However, when $t_{Ag} \sim
%t_{Ar}$ the collisions cause the anisotropy to decay and the maximum anisotropy
%is reduced. Once $t_{Ag} \gg t_{Ar}$, the maximum anisotropy is negligible,
%effectively suppressing the instability.

%Technically, the growth rate is not completely suppressed unless $\nu_e d_e/
%v_T \sim 1$. For  $t_{Ag} \sim t_{Ar}$, the temperature gradient causes the
%anisotropy to grow as $A \sim (t v_T/L_T)^2$, so $1/A dA/dt = 2/t$. This growth
%balances the collisional decay at the anisotropy relaxation time $t \sim
%t_{Ar}$. This balance causes the Weibel instability to grow according to the
%weak anisotropy $A \sim (t_{Ar} v_T/L_T)^2$, which effectively suppresses the
%instability.

\begin{figure}[ht!]
  \noindent\includegraphics[width=3.0in]{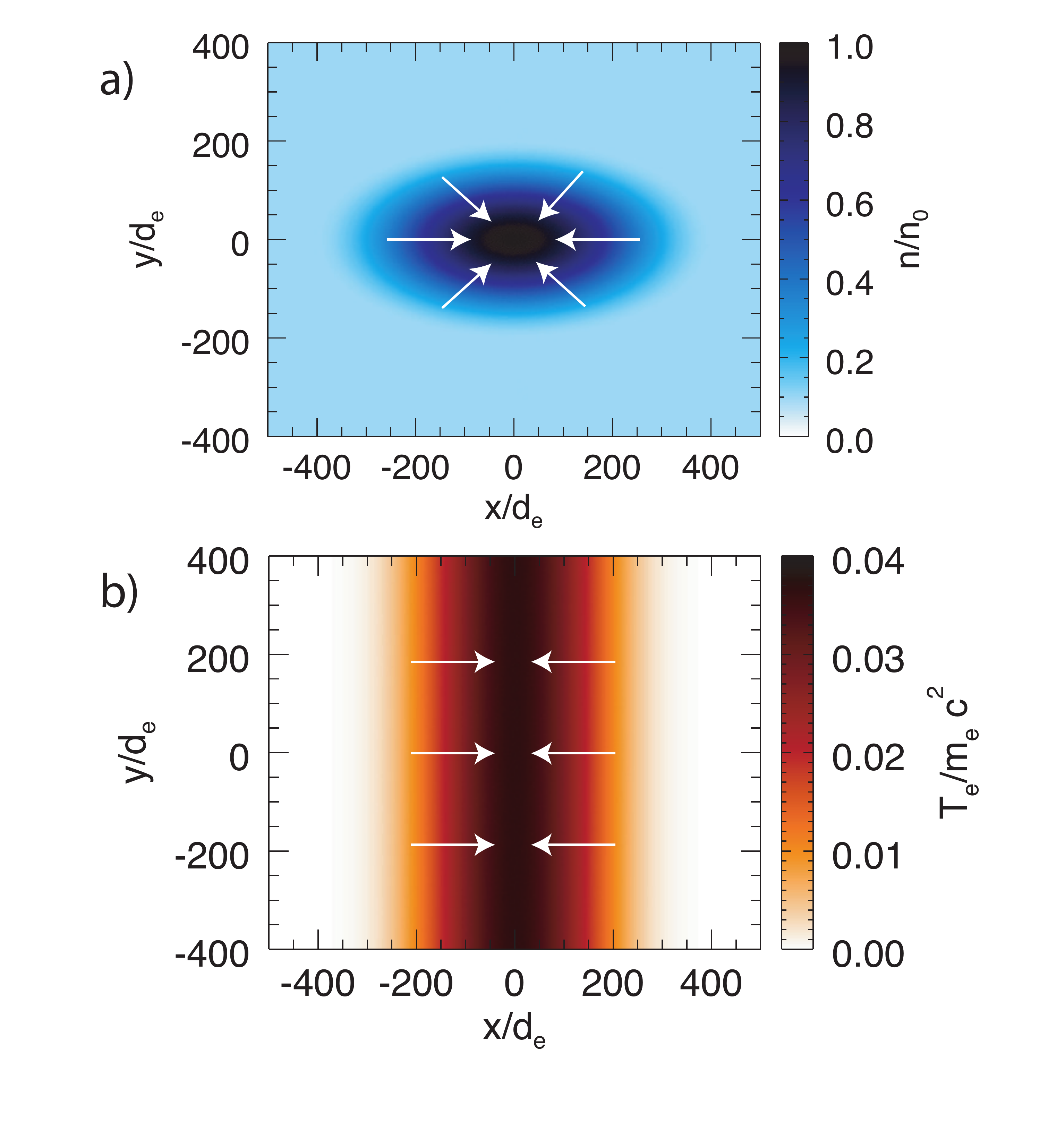}
  \caption{\label{initnT}
	Map of the initial (a) density $n$ and (b) electron temperature $T_e$.
	The gradients are highlighted in white.
}
\end{figure}
\section{Simulation setup}
%Show the Biermann Battery setup, where the temperature is self-consistently generated.
In order to verify and quantify the scalings of \eq{Weibelsuppression},
we performed several simulations of an expanding bubble of plasma using the
OSIRIS framework~\cite{Fonseca2002, Fonseca2008}, while including collisional
effects~\citep{Fiuza2011,Peano2009}, and varying the collisionality $\nu_0 L_T/v_T$ (via
the density).  The
bubble has a peak density $n_0$ and expands into a background $n_b = 0.1 n_0$,
with a peak initial temperature at the center of the box $T_{e0}/m_e c^2= 0.04$
varying only along the $x$ direction to a background temperature $T_{eb} =
0.0025 T_{e0}$, and a realistic mass ratio $m_i/m_e = 1836$, the same as the simulations in
Ref.~\cite{Schoeffler2016} with
\begin{equation}
\label{distributionfunction}
\begin{array} {l}
n = \begin{cases} (n_0-n_b)
\cos(\pi r/2L_T)^2 + n_b, & \mbox{if } r < L_T, \\
n_b, & \mbox{otherwise}, \end{cases} \\ \\
	v_{Te} = \begin{cases} (v_{Te0}-v_{Teb}) \cos(\pi |x|/2L_T)^2 + v_{Teb}, & \mbox{if } |x| < L_T, \\
v_{Teb}, & \mbox{otherwise}, \end{cases} \\
\end{array}
\end{equation}
\be
\mbox{where } r = \sqrt{x^2+{(L_T/L_n y)}^2}, \nonumber
\ee
and $L_n = L_T/2$.
$n_0$ is the reference density used to define
$\omega_{pe}$ and $d_e$ and is used along with the reference temperature
$T_{e0}$ to calculate $v_T$ and $\nu_0$. 
The distributions of density and temperature are shown in \fig{initnT} with the gradients highlighted.
Unless otherwise specified, each
simulation uses $198$ particles per cell on a $12000 \times 12000$ grid
($1500.0 \times 1500.0~d_e^2$). The simulations are run for
$1800.0~\omega_{pe}^{-1}$, with a timestep $dt = 0.07 ~\omega_{pe}^{-1}$.  The
Coulomb logarithm $\ln\Lambda_C$ is calculated automatically depending on the
local parameters.  Only the collisions between electrons and ions are included.
There is an 8 point average over the magnetic fields generated.

In realistic experimental setups, where collisions become important, the
temperatures are lower and the system sizes are larger than we simulate here.
As these parameters are more computationally expensive, we instead vary the
density, allowing for a scaling to realistic parameters. Note that for the
parameters that we simulate, $\ln\Lambda_C$ varies significantly due to the
small values if $\Lambda_C$.  Once $v_T/c$ exceeds $2 \alpha$ ($T_e > 108$ eV),
where $\alpha$ is the fine structure constant, $\Lambda_C$ grows more slowly
with respect to temperature.  Therefore, for a given collision rate,
$\Lambda_C$ is smaller for higher temperatures.  Furthermore, for larger system
sizes, equal collisionalities occur at smaller collision rates, and thus at
larger $\Lambda_C$.

In our simulations the velocity distribution does not necessarily remain
bi-Maxwellian, and we measure the anisotropy using the temperature tensor
$T_{ij} \equiv \int (u_i u_j/\gamma) f(u)/ \int f(u)$ calculated in the species rest
frame. $u_i$ is the proper velocity, $\gamma = \sqrt{1+u^2}$, and $f(u)$ is the
velocity distribution function. $T_{e\parallel}$ and $T_{e\perp} <
T_{e\parallel}$ are eigenvalues of the temperature tensor. We only consider the
in-plane temperatures and assume the out-of-plane temperature is also
$T_{e\perp}$, which has been verified for our simulations to be a reasonable
assumption.

\section{Simulation results}
\begin{figure}[ht!]
  \noindent\includegraphics[width=3.0in]{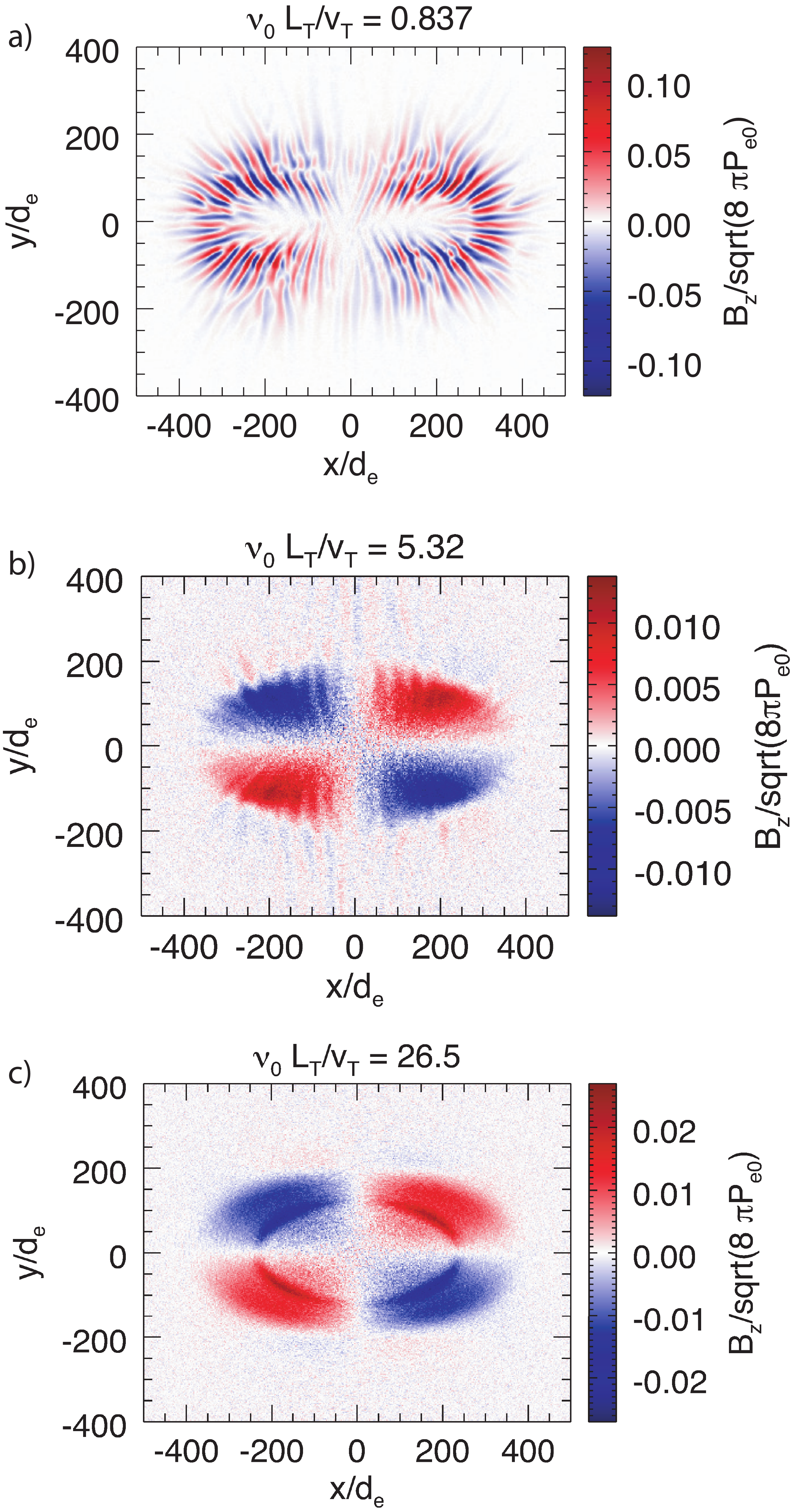}
  %\noindent\includegraphics[width=3.0in]{Bzn24.eps}
  %\noindent\includegraphics[width=3.0in]{Bzn26.eps}
  %\noindent\includegraphics[width=3.0in]{Bzn28.eps}
  \caption{\label{Bz}
	Map of magnetic field $B_z$ at $t\omega_{pe} = 1797.6$. The top panel
shows the simulation with a collisionality $\nu_0 L_T/v_T = 0.837$, where the Weibel instability still exists, but
grows slower and saturates at a lower intensity field. The middle panel
shows a simulation with $\nu_0 L_T/v_T = 5.32$, where the Weibel instability is significantly damped,
and the Biermann field is visible.  The bottom panel
shows a simulation with $\nu_0 L_T/v_T = 26.5$, where there are no traces of the Weibel instability and
a pileup of magnetic flux dragged by the Nernst effect is present.  
}
\end{figure}

In \fig{Bz}, the out-of-plane magnetic field $B_z$ for three
representative collisionalities is presented. The first case (\fig{Bz}a) is
like the previous collisionless studies, where the Weibel magnetic field
dominates compared to the fields due to the Biermann battery.  Only the growth
rate and the strength of the saturated field are modified by the collisions. The second case
(\fig{Bz}b) is the transition scale where the Weibel fields are suppressed,
but still visible, and the Biermann field is the dominant field.  In the third
case (\fig{Bz}c), only the most energetic electrons remain collisionless,
leading to pile-up of Biermann generated fields via the Nernst
effect~\cite{Nishiguchi1984,Kho1985,Ridgers2008,Willingale2010,Willingale2010b,Joglekar2014,Joglekar2016}.
To the best of our knowledge, this is the first time the Nernst effect has been demonstrated
using PIC simulations.

%Note the scaling, the collision rate can be modified by a factor that includes the Z of the
%ions in the plasma, and takes into account the temperature predicted by the experiment.
We can make a prediction where to expect the electron Weibel instability,
depending on the gradient length scale, temperature, density, and charge state
of the ions.  The transition occurs when $\nu_e L_T/v_T \approx 1$, as predicted from \eq{Weibelsuppression}. The three
cases in \fig{Bz} occur at $\nu_0 L_T/v_T = 0.837$, $5.32$, and $26.5$
respectively.  We calculate the local $\nu_A$ at the location where the
instability occurs in the collisionless case ($x/d_e,y/d_e = 150,100$) by
measuring the parameters averaged within a box of $20 \times 10 d_e^2$ at time
$t_c \omega_{pe}= 907$ when the measured growth rate reaches its maximum.
Here, $T_{e\perp,loc}/m_ec^2 = 0.0244$, $A=0.56$, $n_{loc}/n_0 =
0.385$, and the gradient length scale $L_{T,loc}/L_T = 0.2563$.
We thus find the respective cases occur at $\nu_{A,loc} L_{T,loc}/v_{T,loc} =
2.09$, $13.5$, and $63.7$. We therefore confirm that the Weibel instability is
suppressed when the anisotropy relaxation time is smaller than the generation
time ($t_{Ar} > t_{Ag}$), but it is not completely suppressed until $t_{Ar}
\gg t_{Ag}$.

The transition to a regime where no Weibel exists occurs in the simulation with
$\nu_0 L_T/v_T = 5.32$ (\fig{Bz}b). 
%This simulation has no electron-electron
%collisions, $Z=1$, $L_T = 213 nm$, $ln \Lambda_C = 4$, and $v_T/c=0.2\text{
%}(T_e=20.4~keV)$. 
This simulation has no electron-electron
collisions, $Z=1$, and the local $L_{T,loc} = 55 nm$, $ln \Lambda_C = 4.0$, and $v_{T,loc}/c=0.170\text{
}(T_{e,loc}=14.8~keV)$. 
Using this numerical value of $\nu_0 L_T/v_T$ and the scaling from \eq{Weibelsuppression},
our simulation result can be scaled to more experimentally relevant
densities and temperatures, and the general transition density can be expressed
in an engineering formula.
%Engineering form of transition rate.
%\begin{eqnarray}
%	n = 10^{26} \text{cm}^{-3} && \left(\frac{T_e}{20.4~keV}\right)^{2}\left(\frac{1+\sqrt{2}Z}{\sqrt{2}}\right)^{-1}\nonumber\\
%	&& \left(\frac{L_T}{213~nm} \right)^{-1}\left(\frac{\ln \Lambda_C}{4.0}\right)^{-1}
%\end{eqnarray}
%or
%\begin{eqnarray}
%	n_{tr} = 2.89 \times 10^{22} \text{cm}^{-3} && \left(\frac{T_e}{1.0~keV}\right)^{2}\left(1+\sqrt{2}Z\right)^{-1}\nonumber\\
%	&& \left(\frac{L_T}{1.0~\mu m} \right)^{-1}\left(\frac{\ln \Lambda_C}{10.0}\right)^{-1}
%\end{eqnarray}
\begin{eqnarray}
	n_{tr,loc} = 5.42 \times 10^{21} \text{cm}^{-3} && \left(\frac{T_{e,loc}}{1.0~keV}\right)^{2}\left(1+\sqrt{2}Z\right)^{-1}\nonumber\\
	&& \left(\frac{L_{T,loc}}{1.0~\mu m} \right)^{-1}\left(\frac{\ln \Lambda_C}{10.0}\right)^{-1}
\end{eqnarray}
The other two cases correspond to a density $n_{loc} = 0.1 n_{tr,loc}$ (\fig{Bz}a) and $10  n_{tr,loc}$ (\fig{Bz}c).
%1.08e21 and 9.62e20 are the Nernst scaling for the parameters of Fox 2018
%which correspond to 5e20 and 4.5e21 densities

\begin{table}[ht!]
        \centering
	\caption{Measured growth rate $\gamma_m$ and parameters determining the theoretical growth rate $\gamma_t$ at the location where the instability occurs
	($x/d_e,y/d_e = 150,100$) averaged within a box of $20 \times 10 d_e^2$ at time $t_c$ where the measured growth rate reaches its maximum.
	The local density is $0.385~n_0$. For $\nu_0 L_T/v_T = 5.32$, the measured growth rate can be considered $0$.}
	\begin{tabular}{|p{1.2cm}|p{1.5cm}||p{1.0cm}| p{1.0cm}| p{0.8cm}|p{1.4cm}| p{1.0cm}|}
                \hline
		\bf $\nu_0 L_T/v_T$ & \bf $\nu_0/\omega_{pe}$ & \bf $t_c \omega_{pe}$ & \bf $\gamma_m/\omega_{pe}$ & \bf $A$ & \bf $T_{e\perp}/m_ec^2$ & \bf $\gamma_t/\omega_{pe}$ \\ \hline
                $0.00175$ & $8.8 \times 10^{-7}$ & 907.0  & 0.0098  &  0.56  &  0.0244   & 0.0095 \\ \hline
		$0.0145$ & $7.2 \times 10^{-6}$ & 900.6  & 0.0111  &  0.56  &  0.0245   & 0.0096 \\ \hline
                $0.114$ & $5.7 \times 10^{-5}$ & 946.5  & 0.0086  &  0.57  &  0.0241   & 0.0097 \\ \hline
                $0.837$ & $4.2 \times 10^{-4}$ & 1047.0 & 0.0065  &  0.47  &  0.0225   & 0.0072 \\ \hline
		$5.32$ & $2.7 \times 10^{-3}$ & 600.0  & 0.0001  &  0.02  &  0.0224   & 0.0000 \\ \hline
                $26.5$ & $1.3 \times 10^{-2}$ & 500.0  & 0.0001  &  0.02  &  0.0217   & -0.000 \\ \hline
        \end{tabular}
	\label{measuredgrowthrates}
\end{table}

\begin{figure}
  \noindent\includegraphics[width=3.0in]{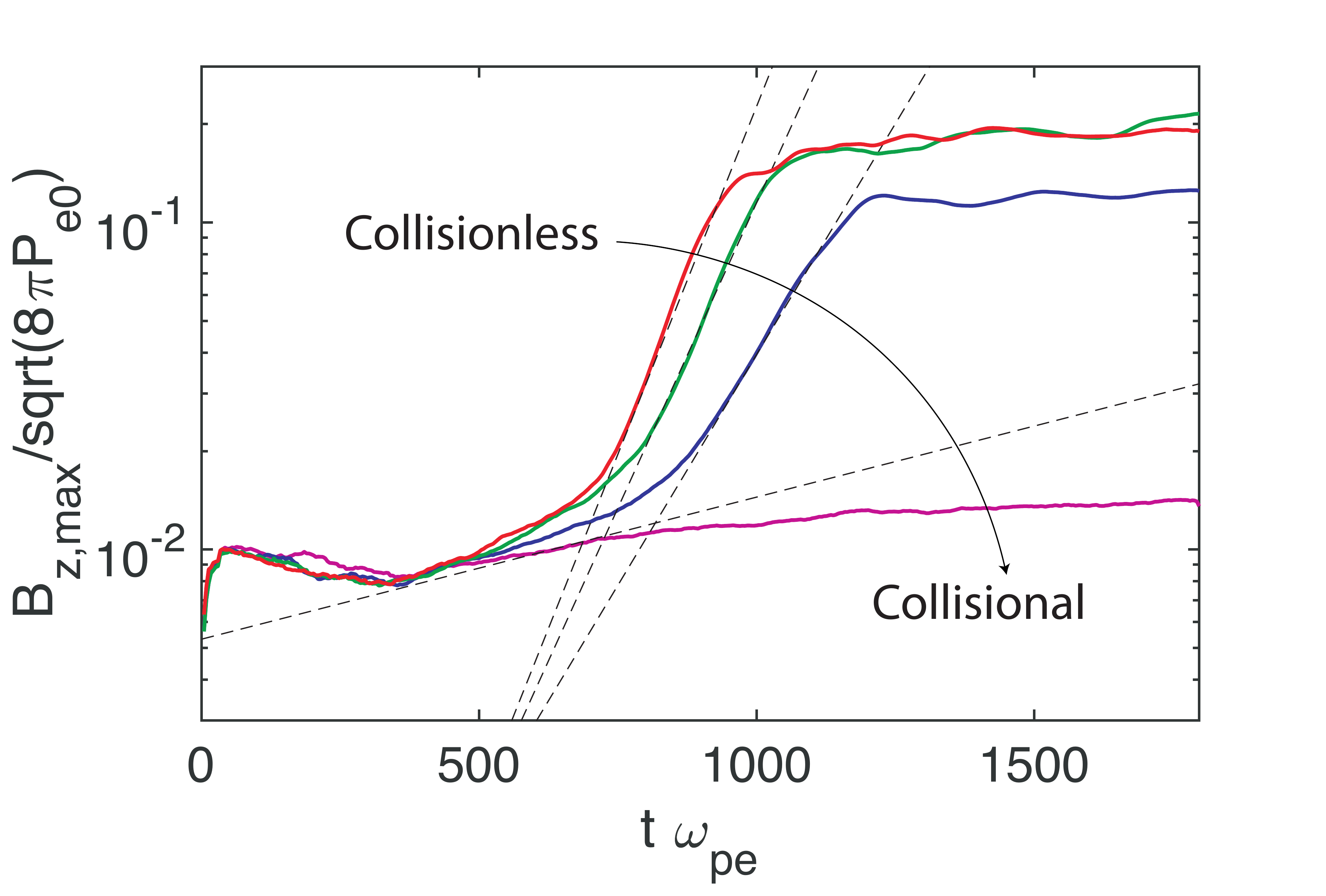}
  \caption{\label{Bmaxev}
	Evolution of the maximum magnetic field $B_{z,max}$ (produced via the Weibel instability) vs. time for simulations with various collisionalities $\nu_0 L_T/v_T = 0.00175$ (red), $0.114$ (green), $0.837$ (blue), and $5.32$ (magneta).
	The measured slopes (at $t_c$ where the slope reaches its maximum) occurring at
	$t_c \omega_{pe} = 907, 946.5, 1047,$ and $600$ are shown in dashed black.
	}
\end{figure}

We show the effects of collisions on the Weibel growth in~\fig{Bmaxev},
plotting the evolution of the maximum magnetic field and exponential fits of
the growth rate. There is a significant change in growth rate
between the essentially collisionless case at  $\nu_0 L_T/v_T=0.00175$, and
$\nu_0 L_T/v_T=0.837$ ($0.0098$ to $0.0065\omega_{pe}$). The growth rate
effectively goes to zero in the case with $\nu_0 L_T/v_T=5.32$.  The measured
growth rates of the magnetic field for each simulation are reported in
Table~\ref{measuredgrowthrates}, along with the local parameters used to
calculate the theoretical Weibel growth rate.
Table~\ref{theoreticalgrowthrates} shows the measured wavenumber of the
instability, the predicted fastest growing wavenumbers, and the growth rates
calculated using these wavenumbers, providing evidence that the observed
filaments are due to Weibel instability.

\begin{figure}
  \noindent\includegraphics[width=3.0in]{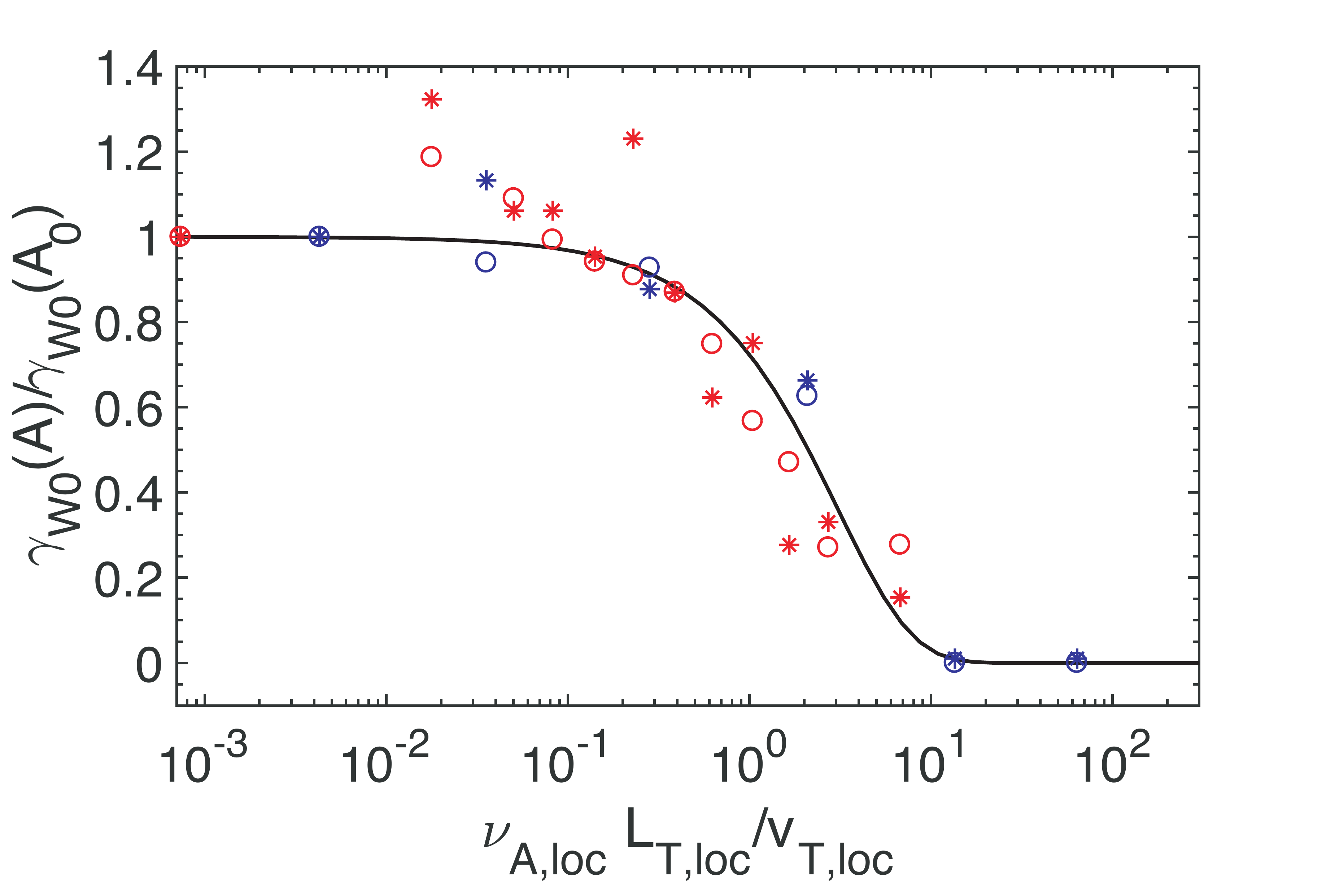}
  \caption{\label{gammavsnu}
	Measured growth rate of the Weibel instability normalized to the
	growth rate measured in the collisionless case for a range of
	simulations with different collisionalities. Simulations with $L_T/d_e
	= 400$  (200) are indicated in blue (red). Circles are calculated from
	magnetic field energy, and stars from the maximum magnetic field.  A
	theoretical estimate of the growth rate is plotted in black using the
	anisotropy from \eq{Avsnu}.
	}
\end{figure}

%\begin{table*}[ht!]
%        \centering
%        \caption{The }
%	\begin{tabular}{ |p{1.3cm}|p{1.7cm}||p{1.0cm}| p{1.0cm}| p{1.1cm}|p{0.8cm}|p{1.0cm}| p{1.3cm}|}
%                \hline
%		\bf n [$cm^{-3}$] & \bf $\nu_c/\omega_{pe}$ & \bf $\bar{\gamma}_m$ & \bf $B_{b0}$ & \bf $\bar{B}_0 e^{\bar{\gamma}_m  t_c}$ & \bf $k d_e$ & \bf $k_{max} d_e$ & \bf $\gamma_t(k)/\omega_{pe}$ \\ \hline
%		$1 \times 10^{18}$   & 0                        & 0.0096  & 0.0038 &  0.015  &  0.281 & 0.2523 & 0.0083 \\ \hline
%		$1 \times 10^{20}$   & $1.9 \times 10^{-5}$   & 0.0090  & 0.0037 &  0.014  &  0.204 & 0.2523 & 0.0081 \\ \hline
%		$1 \times 10^{22}$   & $1.6 \times 10^{-4}$   & 0.0090  & 0.0038 &  0.015  &  0.200 & 0.2546 & 0.0081 \\ \hline
%		$1 \times 10^{24}$   & $1.3 \times 10^{-3}$   & 0.0051  & 0.0037 &  0.013  &  0.208 & 0.2312 & 0.0063 \\ \hline
%		$1 \times 10^{26}$   & $9.7 \times 10^{-3}$   & 0.0     &        &  N/A    &  N/A   & N/A    & 0.0    \\ \hline
%		$1 \times 10^{28}$   & $6.6 \times 10^{-2}$   & 0.0     &        &  N/A    &  N/A   & N/A    & 0.0    \\ \hline
%        \end{tabular}\label{1}
%\end{table*}
\begin{table}[ht!]
        \centering
	\caption{The measured wavenumbers $k$, theoretical fastest growing mode $k_{max}$, and the theoretical growth rates $\gamma_t$ 
	given these wavenumbers.}
	\begin{tabular}{ |p{1.4cm}||p{0.8cm}|p{1.0cm}| p{1.3cm}|p{1.0cm}|}
                \hline
		\bf $\nu_0 L_T/v_T$ & \bf $k d_e$ & \bf $k_{max} d_e$ & \bf $\gamma_t(k)/\omega_{pe}$ & \bf $\gamma_t/\omega_{pe}$ \\ \hline
		$0.00175$   &  0.281 & 0.2523 & 0.0094 & 0.0095 \\ \hline
		$0.0145$   &  0.204 & 0.2523 & 0.0091 & 0.0096 \\ \hline
		$0.114$   &  0.200 & 0.2546 & 0.0091 & 0.0097 \\ \hline
		$0.837$   &  0.208 & 0.2312 & 0.0071 & 0.0072 \\ \hline
		$5.32$   &  0.080 & 0.0502 & -0.000 & 0.0000 \\ \hline
		$26.5$   &  N/A   & 0.0502 & N/A    & -0.000 \\ \hline
        \end{tabular}
	\label{theoreticalgrowthrates}
\end{table}
%more precise nu 1.886  1.580  1.275  9.701  6.649
%Show the growth rates of magnetic fields for the Biermann Battery setup.

The growth rate of the Weibel instability depends on $A$, which depends on the
collisionality as collisions inhibit the anisotropy growth. Due to the
exponential decay of $A$ predicted by \eq{Adecayrate} for a constant $\nu_A$, a good
approximation of the $A$ dependence on the local collisionality is:
\be
\label{Avsnu}
A \approx A_0 \exp\left(-\frac{\nu_{A,loc}L_{T,loc}}{4v_{T,loc}}\right).
\ee

In addition to the simulations presented so far with $L_T/d_e = 400$, we have
simulated several more simulations with $L_T/d_e = 200$ (half the system size,
with constant resolution), where we have also measured the growth rate. In
\fig{gammavsnu} the measured growth rates normalized to the collisionless
growth rates are presented as a function of the local collisionality
$\nu_{A,loc}L_{T,loc}/v_{T,loc}$.  $\nu_{A,loc}$ is calculated as previously
assuming a constant anisotropy $A_0=0.56$ and perpendicular temperature
$T_{e\perp,loc}/m_e c^2=0.0244$.  A theoretical prediction for the growth rate
is given by the Weibel growth rate using the anisotropy from \eq{Avsnu} (black
curve in \fig{gammavsnu}), which agrees with the measured results.
\fig{gammavsnu} also gives evidence that this scaling with collisionality is
independent of $L_T$ for constant $\nu_0L_T/v_T$.

\section{Conclusion}
Using particle-in-cell simulations, we have placed a limit where collisions
will inhibit the generation of the electron Weibel instability in the expansion
of a hot plasma, when $\nu_e L_T/v_T \sim 1$.  While in ~\citep{Schoeffler2014}
it was shown that magnetic fields from the Weibel instability will be larger
than the Biermann field for $L_T/d_e > 100$, we now show this additional limit
due to collisions, where the Biermann field again dominates.

Although the simulations presented here are all 2D, the results
should not differ greatly in 3D. For the collisionless case a 3D simulation
showed similar results for $L_T/d_e = 50$~\citep{Schoeffler2016}. For larger
system sizes we expect Weibel filaments with wavenumbers also out of the 2D
simulation plane, but besides that, the results should remain similar to 2D.

We do not observe the thermomagnetic instability, confirming
Ref.~\cite{Sherlock2020}, but we also do not observe any of the
predicted reduction of the Biermann battery growth.  This is likely because we
start from a Maxwellian distribution, where the Biermann battery should grow
rather than evolve to a such a state by plasma heating and expansion.

This still remains a simplified model, and assumes that the laser interaction
will generate these temperature gradients on a quick enough time scale that
this model is valid. The effects of the laser magnetic fields and heating
processes often occur at the same time as the Biermann and Weibel magnetic
fields grow. It has been shown that for an intense short pulse laser, where the
plasma becomes relativistically hot, the Weibel field can be
observed~\citep{Shukla2019}. 
%I found, however, in unpublished work that for a
%weaker laser, the Weibel instability is not observed.

\paragraph{Acknowledgments.}
This work was supported by the European Research Council (ERC-2015-AdG Grant
No. 695088).

\end{document}